Decision Theory is a Red Herring for the Many Worlds Interpretation

Dr. Jacques Mallah (jackmallah@yahoo.com), 8/18/2008; revised 12/22/11
http://onqm.blogspot.com

Abstract:

There have been many attempts over the years to derive the Born Rule from the wave equation since Everett's Many Worlds Interpretation was proposed; however, none of these have been satisfactory, as shown when critics pointed out loopholes and unsupported assumptions.  In this paper the case is made that the currently fashionable decision-theoretic approach founded by David Deutsch and explicated by David Wallace is likewise unsatisfactory.  A more fundamental computationalist approach is advocated.

Introduction:

Derivation of the Born Rule is the key to showing that the Everett-style Many Worlds Interpretation (MWI) of quantum mechanics, with no hidden variables or modifications to the wave equation, is viable.  Given that the MWI is the simplest interpretation of QM, an uncontroversial derivation of the Born Rule from the wave equation should establish the MWI as the favored interpretation.  Conversely, if the Born Rule can not be so derived, it would be unlikely for the MWI to ever be widely accepted in its unmodified form; and if some other wrong rule is derived from the MWI instead of the Born Rule, it would falsify the unmodified MWI.

There have been many attempts over the years to derive the Born Rule from the wave equation since the MWI was proposed; however, none of these have been satisfactory as shown when critics pointed out loopholes and unsupported assumptions [Kent, Van Esch].  In this paper, I argue that the currently fashionable decision-theoretic approach founded by David Deutsch and explicated by David Wallace [Wallace] is likewise unsatisfactory.

I will then sketch another approach that may offer the best chance of dealing with the issue of how the Born Rule is related to the MWI [Mallah].  This approach frames the issue of probabilities in the MWI in a philosophically fundamental way regardless of whether the outcome is favorable to the MWI or not.

Likelihoods and the MWI:

It is worth mentioning at the outset that there is a misleading connotation in the language of 'likelihood' as used in decision theory regarding the MWI, which is that such a 'likelihood' is often thought of as constant in time, although it need not be.  In ordinary (non-MWI) probabilistic situations, in which a single outcome is chosen from a set of possibilities, the chosen outcome thereafter remains the only actual outcome.  Therefore, in a single-world case it would make no sense to speak of probabilities changing after the fact.

In the MWI, however, all outcomes actually occur. The standard MWI is deterministic, so the "probabilities" are not chances for random outcomes, which would require some non-deterministic element. Whatever physical things the "probabilities" are proportional to could change as a function of time differently within each branch.

And unless some form of tracking of observers (meaning that in principle we could identify something within the wavefunction dynamics as single observer, and then later identify something as the single observer that *deterministically* resulted from the previous observer) is involved, then the "probabilities" for future events in the MWI are not related to incomplete information being available prior to the branching, assuming that the relevant facts about the future behavior of the universal wavefunction are known in advance. In that case the MWI "probabilities" for future events are not probabilities in the sense of subjective uncertainty, either.

The "probabilities" in the MWI are best understood as "effective probabilities": that is, they play the role of probabilities for all practical purposes. These practical purposes include making predictions and retrodictions, and making decisions.

At first sight, that would seem to suggest that the decision-theoretic approach is on the right track. As will be seen, however, it fails to provide any justification for the Born Rule. That doesn't mean that applying decision theory to the MWI is not necessary; it merely means that doing so is not a route to the Born Rule.

Explaining what exactly *would* play the role of 'probabilities' in the MWI is *not* necessary for the purpose of refuting the decision-theoretic claim to derive to derive the Born Rule as done in the next sections. That said, the following considerations address the question of what would play the role of 'probabilities':

Subjective uncertainty *can* come into play in the MWI when considering situations where an observer does not yet know his current location among branches. It is clear that since these kinds of 'probabilities' are those that an observer would expect to use to make predictions about which branch he's on *if he is typical*, the 'amount of consciousness' in each branch is what must be proportional to these 'probabilities'. I call this amount *measure*. The probability of seeing a given outcome is equal to the measure of conscious observations of that outcome divided by the total measure. As shown by the example of death, total measure is not necessarily conserved as a function of time.

Furthermore, when one does correctly apply decision theory to the MWI, there must be no introduction of subjective uncertainties where they *don't* belong. This means that – unless observer-tracking within the wavefunction is involved, and assuming he knows the relevant facts about what the wavefunction will be for each of his possible choices - whatever plays the role of probabilities *for decision-making* (influencing the future) will do so because of the decision-maker's preferences, *not* because of his uncertainties. For example, if the Born Rule holds, he must have a reason to prefer that better outcomes occur in branches of high amplitude while the worse outcomes occur in branches of low amplitude.

It is clear in the 'amounts of consciousness' picture *why* a decision-maker should have unequal preferences (not related to observer-tracking or to ordinary single-world concerns like quality-of-life) among 'branches': the branches with a greater amount of consciousness in them – which is to say, with more people in them – are more important because the good of the many outweighs the good of the few.

It is **not** sufficient to take only probabilities into account, rather than measure, because more conscious life is generally considered desirable; decreasing measure has the same effect as decreasing the number of people. (Failure to account for decreases in measure leads to the "quantum immortality" fallacy.) More details on this and other aspects of how measure leads to "effective probabilities" are discussed in reference [Mallah 2].

If the measures and probabilities change in a way that is for practical purposes unpredictable, a rational decision maker would have nothing to base decisions on. This is a logically possible situation as a consequence of a given ontology. It would certainly not be justifiable to assume that the Everett ontology must be conducive to decision making as a way of 'deriving' the Born Rule from it – and as will be shown, such an assumption would not suffice in any case.

The Decision-Theoretic Approach to the Born Rule:

The decision-theoretic approaches discussed here are based on the attempted derivations of the Born Rule by Wallace [1,2]. The 2005 preprint by Wallace was, at least until his more recent paper, considered by advocates of the decision-theoretic approach to be the most complete and correct presentation of the approach. Recently, in June 2009, Wallace posted a new preprint containing a "formal proof of the Born rule from decision-theoretic assumptions". I will first discuss Wallace's 2005 preprint, and then his 2009 'formal proof'.

Wallace's 2005 preprint:
Decision theory itself can be seen to play a relatively small part in the overall 'derivation' presented in Wallace's 2005 paper, although the 'derivation' is framed in terms of it. The 'derivation' can be divided into the following parts:

1) A 'likelihood' function plays the role for the decision-maker that 'probabilities' would normally play, in assigning a weight to each outcome. The decision-theoretic assumptions employed by Wallace relate to this 'likelihood' and are as follows:

Transitivity: 'If A is at least as likely than B and B is at least as likely than C, then A is at least as likely than C.'

Separation: 'There is some outcome that is not impossible.'

Dominance: 'An event doesn't get less likely just because more outcomes are added to it; it gets more likely iff the outcomes which are added are not themselves certain not to happen.'

These three assumptions are not problematic.

Indeed, corresponding assumptions would certainly hold for probabilities of seeing outcomes, understood as measure ratios, with the rare exception of #2 in the case where the observer is certain to die regardless of the measurement outcome.

2) In order to 'derive' the Born Rule, it is of course necessary to add some physics to the decision-theoretic assumptions. This is done with the assumptions of "weight richness" and "Equivalence".

Wallace uses the term 'weight' to mean the squared amplitude of a branch. While this terminology is not by itself an assumption, it is undesirable to use such a connotation-laden term prior to the derivation of what the significance of such a quantity would be.

Weight richness: A set M of quantum measurements is rich provided that, for any positive real numbers $w_1, \ldots, w_n$ that sum to 1, M includes a quantum measurement with n outcomes having squared amplitudes $w_1, \ldots, w_n$.

The 'weight richness' assumption will be used to help generalize the case of equal-amplitude branches. It is an idealization, but is not really problematic.

The 'Equivalence' assumption is what does the real work, and it is the main point of controversy with the approach.

Equivalence: If two events (outcomes) have the same squared amplitude, then they should be assigned equal 'likelihood' by the rational decision maker.

The Born Rule is then derived by noting that if the amplitudes are not equal, then further measurements can (for rational ratios of the squared amplitude) partition the set of events into a large number of outcomes, each of which does have equal amplitude. By 'Equivalence', such unitary transformations don't affect the likelihood of each type of outcome, since the squared amplitude of a transformed state is equal to that of the initial state. The probabilities are then obtained by counting these branches of equal amplitude. The three assumptions of part 1) come into play only to extend this to the case of irrational ratios.

The 'Equivalence' assumption is thus equivalent to the Born Rule. In this sense, it may already be said that the focus on decision theory is a red herring; 'Equivalence' is the only nontrivial part of the whole approach.

The approach could still be worthwhile if 'Equivalence' were somehow more *directly* justifiable than the Born Rule.

3)      Defense of the 'Equivalence' assumption.

Wallace offers a number of attempted justifications for the 'Equivalence' assumption.

a)      'Measurement Neutrality'

This assumption, based on Deutsch's original argument, states that "once we have specified which system is being measured, which state that system is being prepared in, and which observable is being measured on it, then we have specified everything that we need to know for decision-making purposes".

Wallace points out that some measurement processes can be described in more than one way; for example, it is an arbitrary convention to classify a certain part of the process to be part of 'state preparation'. Likewise, what quantity has been 'measured' is often a matter of convention; measuring a quantity might be done by measuring some function of that quantity. In these cases in which the *same physical process* is being described in different ways, it is of course true that the choice of description could not affect the probabilities.

Does this suffice to establish 'Measurement Neutrality'? No, because *different physical processes* could correspond to the same system / preparation / measured observable class. In particular, the way in which the measurement is carried out could be different, such as the measurement of some other (finer-grained) observable at the same time (creating more sub-branches), or any other variation which might affect the probabilities.

Wallace admits that "The reasons why we treat the state/observable description as complete are not independent of the quantum probability rule … such a justification is in danger of circularity in the present context", but for some reason, he does not follow the implication of his own admission. 'Measurement Neutrality' in the general case of different physical processes is not given *any* support by his argument about same physical process cases, which serve only as another red herring.

However, he does follow a different line of argument ("measurements may not be a natural category" from a certain point of view) leading him not to *rely* on 'Measurement Neutrality' as the basis for 'Equivalence' after all. He therefore goes on to present more-direct arguments for 'Equivalence', which I will consider next.

Wallace proceeds by a series of "too quick" arguments, continuing this pattern. That is, he presents an argument and says it establishes his conclusion, and then next he admits that there is a hole in the argument and points out what he thinks that is, then presents an argument to plug the hole, and repeats the process until he is satisfied. The argument fails if there is either a hole he overlooked, or if one of his plugs fails.

b)      Direct argument for 'Equivalence'

The "erasure" argument:

Assume that the rational agent only cares about obtaining some reward, and wants to know his probability of doing so for each of two different "games". The difference between the games is which of two measurement results, with each outcome being of equal amplitude-squared, will give him the reward. 'Equivalence' in this context will be established if he must be indifferent between choosing either game because the outcomes are of equal probability. If instead there is some other criterion which would bear on probabilities, he could prefer one game over the other based on that.

Erasure: If the actual result is 'erased', so that the agent can not learn of it, then the 'only' difference between the outcomes will be the reward. The two games become 'identical', so the agent must be indifferent between them. And he only cares about the reward, so he will not care about whether the erasure occurred or not.

The hole: The games are not identical; the physical state of the outcomes in which he does or doesn't get the reward will be different depending on which game was chosen. Even though the agent doesn't care about the states as such, he does care about the probabilities, which the physical states could determine.

The plug: Wallace admits the problem, and attempts to solve it by again appealing to the agent's indifference to anything but the reward. He notes that "since he is indifferent to the erasure, let alone its details, he does not care which" reward state results from the erasure.

The real hole in the plug: Even though the agent doesn't care about the states as such, he does care about the probabilities, which the physical states could determine.

Wallace's limited version of the hole in the plug: Wallace next admits – seemingly out of the blue since his defense of 'Equivalence' has not yet mentioned branching – that there is a hole in the argument, which is that branching may be relevant.

(Of course, this is just a special case of "the real hole" I mentioned above, an aspect of the physical states that could play a role in determining probabilities. Branching is surely one of the more popular and plausible candidates for such a property of physical states, but is by no means the only possibility. For example, at the "Many Worlds at 50" meeting [MW@50] David Albert discussed a toy example of another such property, namely the "fatness" (mass) of the observer multiplied by the squared amplitude.)

c) Branching Indifference

Wallace proposes to plug the branching hole by establishing "Branching indifference: An agent is rationally compelled to be indifferent about processes whose only consequence is to cause the world to branch, with no rewards or punishments being given to any of his descendants."

He discusses three arguments in support of this conjecture:

His first argument comes down to the statement that "if we divide one possible outcome into equally-valuable suboutcomes, that division is not decision-theoretically relevant." The hole in that statement should be obvious - namely, that if the process has an effect on probabilities, then it certainly will be relevant.

Second (since he admits that some people will reject the first argument) he argues that since branching occurs frequently and due to microscopic events everywhere, it is not possible in practice to take branching into account in decision making. This has many holes: 1) it might be possible to take it into account under certain circumstances, such as by increasing the rate of entropy production, 2) if the best we can do is give up on branch counting and weight every possibility equally, then there would also be no reason to take amplitude-squared into account, and 3) more importantly, the question of whether branching affects probabilities is utterly unrelated to what strategies can be pursued in practice.

Third (since he admits that some people will also reject the second argument) he argues that in quantum mechanics, there is actually no such thing as the number of branches, and therefore branching can not matter. This, finally, is the heart of his argument and it is where he will terminate his cascade of arguments, holes, and plugs.

4)      Deriving the Born Rule by process of elimination?

Wallace makes a number of arguments to the effect that the number of branches is not definable; and as he correctly points out "the models of splitting often considered in discussions of Everett — usually involving two or three discrete splitting events, each producing in turn a smallish number of branches — bear little or no resemblance to the true complexity of realistic, macroscopic quantum systems." My comments will follow each line of argument in turn.

a)      QM models of macroscopic systems are infinite-dimensional.

Presumably this refers to the universe as a whole. While it is true that infinite systems present difficulties for counting procedures, this can be dealt with by use of limiting procedures such as regularization to a finite size that is allowed to grow. Finite size field systems have infinite degrees of freedom but at finite energy can be described with a finite number of them.

b)      In such models the decoherence basis is usually a continuous, rather than discrete basis.

Some symmetry or limiting procedure may apply to overcome this as well. If not this would seem to indicate that decoherence basis would not be a good candidate for counting of branches.

c)      Decoherence is an ongoing, continuous process as opposed to a discrete branching event.

d)      The decoherence basis is not precisely defined, either in terms of coarse-graining or position in Hilbert space.

These points c) and d) are true, and this is a valid ground for stating that the decoherence basis would not work for branch counting. However, Wallace errs when he states that this is "the only available method of 'counting' descendants". I will suggest use of another method below.

e)      Branching occurs very often in many natural processes, not just in identified measurements.

This is true, but irrelevant to the argument that the number of branches could not be defined.

From these arguments, Wallace concludes that there is *not even an approximate way* to define the number of branches. The emphasis on this point, as opposed to a mere lack of a precise way to make the definition, is due to his need to rely on the approximately-defined decoherence basis to identify different outcomes and assign them squared amplitudes; he can not insist on rigorous precision since his own approach lacks it.

Now he pulls a non sequitur: Assuming that branch counting is impossible in the MWI, Wallace concludes that therefore the Born Rule must be true in the MWI since it can be defined without reference to branch counting. But there is no justification for such a conclusion. Rather, *if* branch counting is impossible in the MWI, it may be that it is impossible to derive the the rule for its probabilities. And as mentioned above, branch counting is far from the only possibly other than the Born Rule that does not rely on branch counting (e.g. Albert's amplitude-squared-weighted fatness measure).

Wallace usefully summarizes his arguments as follows:

"Some physical difference between games might be:
1. A change to a given branch which an agent cares about;
2. A change to a given branch which an agent doesn't care about;
3. A change to the relative weights of branches; or
4. A splitting of one branch into many, all of which are qualitatively identical
for the agent's descendants in that branch.

By branching indifference, (4) may be discounted. Any change of type (1)
may be incorporated into an agent's utility function without affecting the probabilities.
Changes of type (2) can be erased — by definition the agent doesn't
care about the erasure. This only leaves changes of type (3), which cannot be
erased on pain of unitarity violation."

In fact his #1 and #2 examples of 'differences between games' already form a complete set of possibilities (with the understanding that multiple branches may be changed in such ways simultaneously). But he then has to go on to consider "other" possibilities because in fact what the agent does or doesn't care about is merely a red herring. Nor is erasure a real issue: If a change affects probabilities but can be erased, it would just mean the agent should care about it (yet incorporating it into the utility function would *not* be the right way to proceed). For example, possibility #4 would surely be an example of either #1 or #2, yet he knows he has to address it separately. So he comes up with possibilities #3 and #4, where the real action is. He argues that #4 is undefined, leaving him with #3.

To recap, there are three clear holes in this argument: First, his list of possibilities (not counting #1 and #2) is not at all exhaustive. Albert's weighted fatness measure could have been #5 on his list. He might argue that it already falls under #1 or #2, but then so must his #4 and that certainly means nothing.

Secondly, if branch counting is impossible in the MWI, it does not follow that a 'branching indifferent' probability rule must be true for the MWI. For example, if the probabilities are proportional to the number of branches, and that number can not be defined in the MWI, then the probabilities can not be defined in the MWI. This would be the case if the MWI is the wrong sort of model to give rise to conscious observers, as the Pilot Wave advocates would have us believe. Thus, a separate argument would be needed to show some plausibility to the idea that probabilities would still *make sense* in the absence of what they intuitively would seem to require, countable branches. Decision theory does not do so; it merely relates to what we should do with the probabilities if we know them.

Finally, he has not truly shown that the strategy of 'counting' things is impossible for the MWI ontology, merely that naïve branch counting is undefined. This suggestion will be expanded on below, after discussing Wallace's 2009 paper.

<u>Wallace's 2009 'A formal proof of the Born rule from decision-theoretic assumptions':</u>

Although impressive at first glance, the title of this paper is actually quite revealing. Wallace is not really proving the Born Rule here, but merely that his assumptions imply it. The real question, of course, is whether we should grant the assumptions.

The assumptions listed in this paper are not the same as those in his 2005 preprint. Notably, the controversial assumption called "Equivalence", which explicitly invokes squared amplitude, is not one of the starting assumptions; instead, it emerges as a lemma. Has Wallace succeeded, this time, in actually deriving the Born Rule without any unjustified assumptions?

A number of new assumptions are made in place of the old ones. The assumptions are divided into three categories: four 'axioms of richness', which concern which acts are available to the agent; two 'general principles of rationality', which constrain the preference order agents can have in either classical or quantum contexts; and four 'rationality axioms specific to the Everettian context'.

Upon reading all of these assumptions, it is not obvious at first which of them (if any) would be problematic. Luckily, Wallace helpfully includes near the end of his paper a discussion of which of his assumptions is violated by various proposed 'alternative' probability rules. He actually discusses Albert's Fatness Measure in this section, and the discussion is quite revealing:

"The fatness rule

Description: each branch is given a probability proportional to its quantum mechanical weight multiplied by the mass of the agent in kilograms (such that the total probability is equal to one). Utility is maximised with respect to this probability.

Origin: David Albert (in conversation, and in his contribution to this volume).

Rationale: Albert says, tongue-in-cheek, that an agent should care about branches where he is fatter because "there is more of him" on that branch. He isn't serious, though: the rule is purely presented as a counter-example.

Why it is irrational: It violates diachronic consistency. Albert's agent is (ex hypothesi) indifferent to dieting. But he is not indifferent to whether his future selves diet: he wants the ones on branches with good outcomes to gain weight, and the ones on branches with bad outcomes to lose weight."

The foolishness of Wallace's argument here seems self-evident: He states that rational action would be impossible if the Fatness Rule were true. Yet it is quite obvious that rational action would be perfectly possible.

If probability were proportional to fatness, the most likely reason would be that measure of consciousness is proportional to fatness. Assuming that, the rational strategy is obviously to increase fatness (though limited by longevity impacts) in worlds with net positive utility, and decrease it (or commit suicide) in worlds with negative utility. Nor would there be any reason that one's desires (utility function) must change as a function of time in order to implement this strategy.

Wallace's argument hangs on what he calls 'diachronic consistency', which is one of his 'general principles of rationality' that is supposed to apply in either classical or quantum contexts. Informally, he describes the requirement as saying that "If his preferences do not remain consistent over this timescale, deliberative action is not possible at all".

Sounds reasonable, right? People's utility functions (that is to say, their desires) don't remain constant but we can grant that our hypothetical decision maker does not alter his utility function over the timescale in question and see if that leads to the Born Rule.

But, as the Fatness Rule example shows, there is much more going on here than a simple assumption of an unchanged set of preferences. Wallace defines it as follows:

**Diachronic consistency:** If U is available at $\Psi$, and (for each i) if in the ith branch after U is performed there are acts $V_i$, $V'_i$ available, and (again for each i) if the agent's future self in the ith branch will prefer $V_i$ to $V'_i$, then the agent prefers performing U followed by the $V_i$s to performing U followed by the $V'_i$s.

In other words, if after U has been performed his future self would prefer $V_i$ to $V_i$', (implying that it has a higher utility per unit measure) then the agent must prefer performing U and then Vi to performing U and then Vi'.

What is the problem with this? Whatever determines the effective probabilities could change as a function of time, and as a function of what acts are performed, as it does in the Fatness Measure example. That would cause the *utilities* of outcome-branches to change *even while his preferences remain constant*. There is thus **no** similarity whatsoever here to the case of a classical observer changing his preferences.

An example is as follows: Bob can inherit $1 million, or he can split into two identical copies of himself each of which inherits $500,000, doubling his measure of consciousness. He *would* prefer having $1 million to having $500,000. According to Wallace's interpretation of quantum 'diachronic consistency', that would mean that he must prefer the first option. But obviously, it is perfectly rational that he might prefer the second option, and most people *should* prefer it unless they think there in no value in increasing the amount of human life.

Wallace goes on to say "This is perhaps a good point to recall the rationale for diachronic consistency: rational action takes place over time and is incompatible with widespread conflict between stages of an agent's life. In the case of the fatness rule, agents have motivation to coerce their future selves — by hiring "minders", say — into dietary programs that they will resist. Multiply this conflict indefinitely many times (for branching is ubiquitous) and rational action becomes impossible."

Is that true? Not if the fatness measure is proportional to the amount of consciousness, since the future selves' desire for more measure in certain branches would be the same as that same desire on the part of their past selves, assuming of course that desires and preference do remain constant. Wallace's mistake in thinking that such a conflict would arise is due to failure to fully appreciate the fact that the "probabilities" in the MWI would change with time if measure is not conserved; thus, a person might not enjoy being fat, but would still make himself stay fat even *after* "splits" have occurred, in order to maintain the "probability" or measure of consciousness of his immediate future self. The disanalogy with classical diachronic consistency is thus complete.

Wallace's quantum 'diachronic consistency' would only make sense if measure of consciousness is conserved as a function of time (or more generally, if whatever plays the role of probability in the MWI is a conserved quantity). And once you help yourself to that as-yet-unjustified assumption, it doesn't take much else to conflate 'probability' with one of the only conserved quantities available in the MWI's configuration space, squared-amplitude; there would be no need for any talk of decision theory just to make that short step. Wallace is, in effect, assuming at the outset what he set out to prove.

That observation leads to another good counterexample to the idea that Wallace's assumptions are required by rationality: that of a universe in which the Hamiltonian is not Hermitian. In such a universe, squared-amplitude need not be conserved; it could change as a function of time in a branch-dependent way. If the change is slow enough, this might even be true of our own universe as far as we can know. According to Wallace's assumptions, rational action *would not be possible* in such a universe, since he would have *nothing* to hang onto as a conserved quantity. But rational action would certainly be possible, given knowledge of the measures; just maximize the time integral of (utility per unit measure multiplied by measure).

Quantum Mechanics meets Functionalist Philosophy of Mind:

The MWI, with the traditional Everett ontology, gives a purely mathematical description of the physical world. Thus it is free of the observer-dependent physics that appears to plague Copenhagen-like interpretations.

However, like any sufficiently complete theory of nature we know that if true it must allow for physics-dependent observers, because we know that we exist. That is, the physical world it describes mathematically (if this is to be a sufficiently complete model) must give rise to observers in a way that depends only on which mathematics it provides for the system. Reductive functionalist philosophy of mind, in which physical systems give rise to observers by virtue of having the right mathematically describable properties (functioning), is thus typically a background assumption for MWI advocates.

In [Brown and Wallace], a case against the Pilot Wave Interpretation (PWI) is made on functionalist grounds, as MWI advocates often have done. The PWI ontology adds 'hidden variable' particles without changing the wavefunction. The basic functionalist argument against it is this: If functionalism is true, then if the wavefunction can give rise to observers in many branches by means of its functioning (needing no 'magic ingredient') it will do so (as in the MWI), even if additional particles were added as per the PWI. There would be a relatively small number of extra observers that depend on the particles. The particles would thus be superfluous, since even if they exist we would likely be wavefunction-based observers in an MWI-style multiverse.

Functionalism is most precisely formulated as computationalism, in which it is noted that the mathematically describable properties that should matter are the computations (which need not be digital) implemented by the system. Computations, in a nutshell, are formal systems in which structured states undergo transitions according to specified rules; in other words, models of structure and function. Implementation of a computation by a physical system (though a precise criterion for it remains controversial [Mallah]) means that the system has characteristics reflecting the structure and function represented by that computation. Actually, there is a slight generalization of the idea of computations – to causal chains – which better characterizes function in some possible situations involving inert components, and will be incorporated into future work.

Computationalism implies that the probabilities (using that term here in the MWI sense of representing the effective fractions of observers seeing various outcomes, not implying any random elements) are determined by the structure and function of the physical system. Since we are assuming a reductive worldview in which there are no special laws of physics that involve consciousness, the precise form of this dependence (while philosophically controversial) would have to be a mathematical fact and cannot depend on the actual nature of physical systems. (The Decision Theory approach does not seem likely to be extensible to the general case.)

Thus, if one knew the correct functional dependence, one could calculate the correct probabilities that would occur, given any mathematically describable system as being the physical one – whether it is a system of classical particles, classical waves, quantum mechanical wavefunctions, discrete bits, or whatever. The correct way to calculate probabilities in Quantum Mechanics must be only a special case of this more general mathematical relationship.

The natural candidate for this relationship between the system and the probabilities is that it be some form of counting of the implementations. Thus, if there are 2 identical computer brains (each implementing the same computation) and 1 different computer brain, the probability for being one of the identical brains will be 2/3.

The Everett-style MWI then stands or falls according to whether proper counting of the implementations would yield the Born Rule. Unfortunately, it is not clear how to properly count the implementations, and there are other issues that could affect such a count such as the initial conditions possibly creating 'noise' in the wavefunction. [Mallah] It would be satisfying enough to show that there is *any* method of counting implementations that would be consistent with the Born Rule given only the Shrodinger equation ontology (or instead the Heisenberg picture ontology) and which does not have other implausible implications.

Assuming computationalism, if a proper evaluation of the general procedure to find probabilities with computationalism, when applied to the Everett-style MWI, does not yield the Born Rule, then the Everett-style MWI must be false. If one does not assume computationalism and tries to hold to the MWI regardless, it begs the question of what should replace computationalism and would suggest that the Everett ontology is incomplete.

However, that should be of little comfort to the advocates of most other interpretations. As explained above, adding Pilot Wave particles would not in fact alter the probabilities since the wave function implementations would still dominate. The exception would be if a very large or continuous set of many Pilot Wave particle sets is added, creating a Many Worlds Interpretation with hidden variables. Another plausible possibility would be that none of the existing interpretations of QM is correct and that only by using the full theory that would correctly incorporate quantum gravity (perhaps having a low energy limit resembling a many worlds hidden variables ontology) could the Born Rule be correctly derived from computationalist considerations.